\documentclass[twocolumn,prd,nofootinbib,aps,prl,floats,floatfix,amsmath,amssymb,longbibliography,secnumarabic]{revtex4-1} %
\usepackage[english]{babel}

\usepackage[letterpaper,top=2cm,bottom=2cm,left=3cm,right=3cm,marginparwidth=1.75cm]{geometry}

\usepackage{amsmath}
\usepackage{graphicx}
\usepackage[colorlinks=true,citecolor=red]{hyperref}

\newcommand{\beq}{\begin{equation}}
\newcommand{\eeq}{\end{equation}}

\newcommand{\bea}{\begin{eqnarray}}
\newcommand{\eea}{\end{eqnarray}}

\begin{document}

\title{Comment on ``Neutrino oscillations originate from virtual excitation of $Z$ bosons'' and ``Neutrinos produced from $\beta$ decays of neutrons cannot be in coherent superpositions of different mass eigenstates''}
\author{James M.\ Cline}
\email{jcline@physics.mcgill.ca}
\affiliation{McGill University Department of Physics \& Trottier Space Institute, 3600 Rue University, Montr\'eal, QC, H3A 2T8, Canada}
\affiliation{Niels Bohr International Academy,The Niels Bohr Institute,
Blegdamsvej 17, DK-2100 Copenhagen Ø, Denmark }

\begin{abstract}
It was recently claimed (\url{https://arxiv.org/pdf/2407.00954}) that neutrino oscillations through conventional mass mixing are forbidden by energy conservation, and that they in fact arise from virtual $Z$ boson exchange.
I explain why both claims, and therefore a repetition of the claims in
\url{https://arxiv.org/pdf/2410.03133},
are incorrect.

\end{abstract}

\maketitle

Recently Ref.\ \cite{Zheng:2024wkp} argued that neutrinos are prevented from being in a superposition of different mass eigenstates by energy conservation.
The author, being aware that neutrino oscillations have been observed in experiments, then proposes that flavor oscillations occur as a result of 
virtual $Z$ boson exchange.

To prove the first point, the author assumes that the neutrino in question is in an eigenstate of momentum.   If it is a superposition of different mass eigenstates, then clearly it cannot also be an eigenstate of energy. However the author believes it {\it should} be an eigenstate of energy as well.  He therefore concludes that a quantum superposition of such states is not allowed, and they can only be in a classical superposition, described by a diagonal density matrix. 

In reality, a neutrino emitted in a weak interaction is a flavor state, which is
a superposition of mass eigenstates.  Whether the states in this superposition have exactly the same energy, momentum, or neither, is irrelevant to the fact that they will oscillate.   This age-old issue has been discussed widely in the literature; see for example \cite{akhmedov2009paradoxes,cohen2009disentangling,giunti1991neutrinos,rich1993quantum,akhmedov2017collective,lipkin2004coherent,kayser1981quantum,akhmedov2011neutrino}.

To find some other explanation for neutrino oscillations, the author of Ref.\ \cite{Zheng:2024wkp} has reinvented the concept of a one-loop correction, using cumbersome nonrelativistic perturbation theory, to compute the self-energy correction to neutrinos from virtual $Z$ boson exchange.  In addition he reinvented regularization,  in the form of a Lorentz-violating form factor.  Unfortunately he assumed an incorrect starting point, including tree-level flavor changing neutral-current interactions of $Z$ with the neutrinos.  This of course is not what nature has chosen; therefore the results found in \cite{Zheng:2024wkp} are incorrect.

A talented undergraduate from my particle physics class might have said, ``Oh, maybe he meant to use virtual $W$ exchanges, since those have the flavor structure he assumed.''  Then a talented graduate student from my QFT course could add that this must result in wave function renormalization only, since the neutrino masses are protected by chirality.  The wave function renormalization would have a flavor-diagonal, divergent contribution, and a flavor-changing finite part, proportional to $U_{\alpha i} m_i^2 U^\dagger_{i\beta}$, where $U$ is the PMNS matrix, and $m_i$ are the charged lepton masses.  This could contribute to the running of the neutrino masses and mixings, but is not the origin of flavor oscillations.

In a follow-up paper \cite{zheng2024neutrinos}, these claims were repeated and applied to the case of 
neutrinos coming from neutron beta decays.  But since the first paper is incorrect, this formalism is not valid for beta decays of neutrons, nor tritium, carbon-11, carbon-14, fluorine-20, magnesium-23, potassium-37, holmium-163, bismuth-210 \dots

\bibliographystyle{utphys}
\bibliography{sample}

\end{document}